\documentclass[%
    reprint,
    twocolumn,
    amsmath,
    amssymb,
    aps,
    prstab,
]{revtex4-2}

\usepackage[export]{adjustbox}
\usepackage{graphicx} 
\usepackage{dcolumn}
\usepackage{bm} 
\usepackage[caption=false]{subfig}
\usepackage{appendix}

\newcommand\ddfrac[2]{\frac{\displaystyle #1}{\displaystyle #2}}

\begin{document}

\title{N-dimensional maximum-entropy tomography via particle sampling}

\author{Austin Hoover}%
\email{hooveram@ornl.gov}
\affiliation{Oak Ridge National Laboratory, Oak Ridge, Tennessee 37830, USA}

\date{\today}

\begin{abstract}

We propose a modified maximum-entropy (MENT) algorithm for six-dimensional phase space tomography. The algorithm uses particle sampling and low-dimensional density estimation to approximate large sets of high-dimensional integrals in the original MENT formulation. We implement this approach using Markov Chain Monte Carlo (MCMC) sampling techniques and demonstrate convergence of six-dimensional MENT on both synthetic and measured data.
\end{abstract}

\maketitle

\section{Introduction}

In plasma and beam physics, tomographic methods may be used to infer the distribution of particles in 2D, 4D, or 6D phase space from measured 1D or 2D projections. While 2D tomography is an established technique \cite{Mckee_1995}, 4D and 6D tomography introduce several challenges. The first challenge is to fit the data: searching the space of high-dimensional distribution functions is not straightforward \cite{Wolski_2020, Wolski_2023, Roussel_2024_6d}. The second challenge is to regularize the solution and quantify the reconstruction uncertainty: increasing the phase space dimension can generate an ill-posed inverse problem \cite{Hoover_2024_mentflow}.

Entropy maximization \cite{Caticha_2004, Presse_2013} is well-suited to solve the severely ill-posed inverse problems arising in high-dimensional tomography. The MENT algorithm \cite{Minerbo_1979, Dusaussoy_1991, Mottershead_1996} uses the method of Lagrange multipliers to generate an exact constrained entropy maximum via an unconstrained optimization. While MENT is valid for any number of dimensions and is compatible with arbitrary phase space transformations, its computational cost scales exponentially with the dimension of the phase space and the measurement resolution. While there have been several reported 4D MENT reconstructions \cite{Minerbo_1981, Wong_2022, Hoover_2024_four}, there have been no reported 6D reconstructions.

MENT struggles with high-dimensional problems because of the large number of high-dimensional integrals required to map the reconstructed distribution to its projections. We propose a modified algorithm that uses particle sampling and low-dimensional density estimation to approximate these integrals. We implement this approach using Markov Chain Monte Carlo (MCMC) sampling techniques and demonstrate convergence of 6D MENT for both synthetic and measured data sets.

\section{Background}\label{sec:background}

Our task is to infer a probability distribution $\rho(\bm{x})$, defined over phase space coordinates $\bm{x} \in \mathbb{R}^N$, from its projections onto a set of $M$-dimensional planes. We assume the $k$th of $K$ measurements occurs after a transformation $\mathcal{M}_k: \mathbb{R}^N \rightarrow \mathbb{R}^N$. We measure projections of the transformed coordinates
\begin{equation} \label{eq:forward}
    \bm{u}_k = \mathcal{M}_k(\bm{x})
\end{equation}
onto an axis $\bm{u}_{k_\parallel} \in \mathbb{R}^M$. The projection is the integral of $\rho(\bm{x})$ over the plane $\bm{u}_{k_\perp} \in \mathbb{R}^{N - M}$, which is orthogonal to $\bm{u}_{k_\parallel}$. The distribution must satisfy the following constraints:
\begin{equation} \label{eq:constraints}
    G_k \left[ {\rho}(\bm{x}) \right] 
    = g_k(\bm{u}_{k_\parallel}) - \tilde{g}_k(\bm{u}_{k_\parallel})
    = 0,
\end{equation}
where ${g}_k(\bm{u}_{k_\parallel})$ are the \textit{measured} projections and
\begin{equation} \label{eq:simulated-projections}
\begin{aligned}
    \tilde{g}_k(\bm{u}_{k_\parallel}) 
    =
    \int {\rho} ( \bm{x}(\bm{u}_k) ) d\bm{u}_{k_\perp}
\end{aligned}
\end{equation}
are the \textit{simulated} projections.

The constraints in Eq.~\eqref{eq:constraints} do not identify a unique distribution. To break the degeneracy, we introduce a prior distribution $\rho_*(\bm{x})$, which represents our confidence that a particle would land near $\bm{x}$ if sampled from the true distribution. We update the prior to a posterior $\rho(\bm{x})$ which satisfies the constraints \textit{and} maximizes a functional $S[\rho(\bm{x}), \rho_*(\bm{x})]$, which we call the \textit{entropy}. To determine the entropy functional, we invoke the principle of minimal updating: the prior should be updated as little as possible to fit the data. Enforcing this principle via specific axioms leads to a unique entropy \footnote{This principle of minimal updating is enforced via two axioms which we paraphrase from \cite{Caticha_2021}: (i) \textit{Subset independence}: Probabilities conditioned on one domain should not be affected by information about a different, non-overlapping domain. Consequently, non-overlapping domains contribute additively to the entropy. (ii) \textit{Subsystem independence}: If we assume two variables are unrelated, and if we receive information about each variable separately, then we should not infer any relationship between the variables on the basis of this information. In other words, if we measure the marginal distributions $\{ \rho(\bm{v}), \rho(\bm{w}) \}$, and if the prior does not encode any relationship between $\bm{v}$ and $\bm{w}$, then the posterior should be the product of the marginals: $\rho(\bm{v}, \bm{w}) = \rho(\bm{v}) \rho(\bm{w})$.}: 
\begin{equation}\label{eq:entropy}
    S[\rho(\bm{x}), \rho_*(\bm{x})] =
    -\int{
        \rho(\bm{x}) 
        \log{ 
            \left( \frac{\rho(\bm{x})}{\rho_*(\bm{x})} \right)
        } 
        d\bm{x}.
    }
\end{equation}
The entropy is zero if and only if $\rho(\bm{x}) = \rho_*(\bm{x})$; otherwise it is negative.

The form of the posterior distribution can be derived from the functional
\begin{equation}
    \Psi
    =
    S[\rho(\bm{x}), \rho_*(\bm{x})]
    + 
    \sum_{k}^{} {
        \int
        \lambda_{k}(\bm{u}_{k_\parallel})
        G_k \left[ \rho(\bm{x}) \right]
        d\bm{u}_{k_\parallel}
    },
\end{equation}
where $\lambda_k(\bm{u}_{k_\parallel})$ are Lagrange multiplier functions \cite{Mottershead_1996}. Enforcing zero variation of $\Psi$ with respect to $\rho(\bm{x})$ and $ \lambda_k(\bm{u}_{k_\parallel})$ gives
\begin{equation} \label{eq:ment_solution}
\begin{aligned}
    \rho(\bm{x}) 
    &= 
    \rho_*(\bm{x}) 
    \prod_{k} \exp{ \left( \lambda_k(\bm{u}_{k_\parallel} (\bm{x})) \right) } 
    \\
    &= 
    \rho_*(\bm{x}) 
    \prod_{k} h_k ( \bm{u}_{k_\parallel} (\bm{x}) ).
\end{aligned}
\end{equation}
where we have defined the ``component'' functions $h_k(\bm{u}_k) = \exp(\lambda_k(\bm{u}_k))$.

If we find functions $h_k(\bm{u}_{k_\parallel})$ such that the distribution in \eqref{eq:ment_solution} generates the measured projections, then we have found the constrained entropy maximum. We must solve the system of integral equations that results from substituting Eq.~\eqref{eq:ment_solution} into Eq.~\eqref{eq:constraints}:
\begin{equation}\label{eq:ment_system}
    {g}_k(\bm{u}_{k_\parallel}) = 
    \int 
    \rho_*(\bm{x})
    \prod_{k'} h_{k'} ( \bm{u}_{{k'}_\parallel} )
    d\bm{u}_{k_\perp} .
\end{equation}
Note the hidden dependence of $\bm{x}$ and $\bm{u}_k$ on $\bm{u}_{k'}$ through Eq.~\eqref{eq:forward}. In practice, we evaluate the $h_k(\bm{u}_{k_\parallel})$ on a grid and define continuous functions by interpolation. In this case, the functions are defined by a set of positive numbers $\bm{\theta}$ (the function values on the grid) and we are left with an unconstrained optimization over $\bm{\theta}$.

\section{Methods}\label{sec:methods}

The MENT algorithm solves for the Lagrange multiplier functions using a Gauss-Seidel relaxation method \cite{Minerbo_1979}. Note that $h_k(\bm{u}_{k_\parallel})$ is a function of $\bm{u}_{k_\parallel}$, which is orthogonal to the integration axis $\bm{u}_{k_\perp}$; thus, $h_k$ can be factored out of the integral in Eq.~\eqref{eq:ment_system}. Rearranging gives a set of consistency conditions:
\begin{equation} \label{eq:ment_consistency}
    h_k ( \bm{u}_{k_\parallel} ) =
    \ddfrac
    {
        {g}_k(\bm{u}_{k_\parallel})
    }
    {
        \int 
        \rho_*(\bm{x})
        \prod_{k' \ne k} h_{k'} ( \bm{u}_{k'_\parallel}  )
        d\bm{u}_{k_\perp}
    }.
\end{equation}
MENT begins by initializing the distribution to the prior within the measurement boundaries:
\begin{equation} \label{eq:ment_initialization}
    h_{k}^{(0)}(\bm{u}_{k_\parallel}) = 
    \begin{cases}
        1, & \text{if}\ {g}_k(\bm{u}_{{k}_\parallel}) > 0  \\
        0, & \text{otherwise}
    \end{cases}
\end{equation}
where the superscript represents the iteration number. Next, for each measurement index $k$, $h_k$ is set equal to the right-hand side of Eq.~\eqref{eq:ment_consistency}. The new value is immediately used to update $h_{k + 1}$. In other words, for each iteration $i$ and measurement index $k$,
\begin{widetext}
\begin{equation} \label{eq:ment_update_v1}
    h_k^{(i + 1)} ( \bm{u}_{k_\parallel} ) =
    \ddfrac
    {
        {g}_k(\bm{u}_{k_\parallel})
    }
    {
        \int 
        \rho_*(\bm{x})
        \left(
            \prod_{k' < k} h_{k'}^{(i + 1)} ( \bm{u}_{k'_\parallel} )
        \right)
        \left(
            \prod_{k' > k} h_{k'}^{(i)} ( \bm{u}_{k'_\parallel} )
        \right)
        d\bm{u}_{k_\perp}
    }.
\end{equation}
The iterations may be written in a more interpretable form by dividing Eq.~\eqref{eq:ment_update_v1} by $h_k^{(i)}$:
\begin{equation} \label{eq:ment_update_v2}
\begin{aligned}
    \frac{
        h_k^{(i + 1)} ( \bm{u}_{k_\parallel} )
    }
    {
        h_k^{(i    )} ( \bm{u}_{k_\parallel} )
    }
    &=
    \ddfrac
    {
        {g}_k(\bm{u}_{k_\parallel})
    }
    {
        \int 
        \rho_*(\bm{x})
        \left(
            \prod_{k' < k} h_{k'}^{(i + 1)} ( \bm{u}_{{k'}_\parallel} )
        \right)
        \left(
            \prod_{k' \geq k} h_{k'}^{(i)} ( \bm{u}_{{k'}_\parallel} )
        \right)
        d\bm{u}_{k_\perp}
    }
    .
\end{aligned}
\end{equation}
\end{widetext}
The $h_k^{(i)}$ term in the denominator was absorbed by the integral using Eq.~\eqref{eq:ment_consistency}. Comparing to Eq.~\eqref{eq:ment_system}, we see that the denominator in Eq.~\eqref{eq:ment_update_v2} is the $k$th \textit{simulated projection} (Eq.~\eqref{eq:simulated-projections}) of the MENT distribution function (Eq.~\eqref{eq:ment_solution}) using the up-to-date Lagrange multipliers. Denoting the simulated projection as $\tilde{g}_k(\bm{u}_{k_\parallel})$, Eq.~\eqref{eq:ment_update_v2} becomes
\begin{equation}\label{ment_update_v3}
    h_k^{(i + 1)} ( \bm{u}_{k_\parallel} )
    =
    h_k^{(i    )} ( \bm{u}_{k_\parallel} )
    \left[
    \frac{
        {g}_k(\bm{u}_{k_\parallel})
    }
    {
       \tilde{g}_k(\bm{u}_{k_\parallel})
    }
    \right],
\end{equation}
Finally, we add a learning rate, or damping coefficient, $0 < \omega \le 1$, which slows the journey from the unconstrained to constrained maximum-entropy distribution to stabilize the iterations and handle noisy data \cite{Mottershead_1996}:
\begin{equation} \label{eq:ment_gauss_seidel}
    h_k^{(i + 1)}( \bm{u}_{k_\parallel} ) =
    h_k^{(i    )}( \bm{u}_{k_\parallel} )
    \left[
        1 + \omega
        \left(
            \frac{
                {g}_k(\bm{u}_{k_\parallel})
            }
            {
                \tilde{g}_k(\bm{u}_{k_\parallel})
            }
            - 1
        \right)
    \right].
\end{equation}

Conventional MENT implementations have used numerical integration to simulate the projections $\tilde{g}_k(\bm{u}_{k_\parallel})$ of the distribution function $\rho(\bm{x})$ in Eq.~\eqref{eq:ment_solution}. In this approach, the projected density is approximated by a sum
\begin{equation}
    \tilde{g}_k(\bm{u_{k_\parallel}}) 
    \approx
    \sum_{i}
    \rho(\bm{u}^{(i)}_{k_\perp} | \bm{u}_{k_\parallel})
\end{equation}
over integration points $\{ \bm{u}^{(i)}_{k_\perp} \}$. Let us assume that the integration points form a grid of size $R^{(N - M)}$ and that each projection is defined on a grid of size $R^M$, where $R$ is the grid resolution. Then computing a single projection requires $R^N$ evaluations of the probability density $\rho(\bm{x})$ defined in Eq.~\eqref{eq:ment_solution} ($R^M$ total integrals, each of dimension $R^{(N - M)}$). These integrals can be prohibitively expensive for large phase space dimension $N$ or measurement/integration grid resolution $R$.

We propose to estimate each projection from a batch of particles $\{\bm{x}^{(i)}\}$ sampled from $\rho(\bm{x})$. There are a few potential advantages to this particle sampling approach. First, the phase space transformations are not inverted during the iterations and do not need to be symplectic. This is a convenience when interfacing with particle tracking codes. Second, the computational cost depends only weakly on the measurement resolution. Third, several techniques have been developed to sample from high-dimensional distribution functions without exponential storage requirements. Here, we focus on Markov Chain Monte Carlo (MCMC) sampling techniques \cite{Gelman_1995}, specifically the Metropolis-Hastings (MH) algorithm \cite{Metropolis_1953}. 

The MH algorithm explores the target distribution by a random walk, starting from point $\bm{x}_0$. Given the current point $\bm{x}_t$, a new point $\bm{x}_*$ is proposed from the ``jumping'' distribution $q(\bm{x}_* | \bm{x}_t)$. Given a random number $r \in [0, 1]$, $\bm{x}_t$ is updated as
\begin{equation} \label{eq:metropolis-hastings}
    \bm{x}_{t + 1} = 
    \begin{cases}
        \bm{x}_*  & \text{if}\ r \leq \pi(\bm{x}_* | \bm{x}_t),  \\
        \bm{x}_t & \text{otherwise},
    \end{cases}
\end{equation}
where
\begin{equation}
    \pi(\bm{x}_* | \bm{x}_t) = 
    \text{min} 
    \left(
        1,
        \frac{ \rho(\bm{x}_*) }{ \rho(\bm{x}_t) }
        \frac{ q(\bm{x}_t | \bm{x}_*) }{ q(\bm{x}_* | \bm{x}_t) }
    \right)
\end{equation}
is the acceptance probability. In the long run, the chain is guaranteed to converge to the target distribution $\rho(\bm{x})$. Although there are no guaranteed convergence rates, the algorithm finds wide use in practice. The performance depends primarily on the jumping distribution, which controls the acceptance probability. A common choice is a Gaussian jumping distribution centered on $\bm{x}_t$ with an adjustable covariance matrix. For common convergence diagnostics and tuning strategies, see \cite{Gelman_1995}.

Since MCMC is \textit{gridless}, it has the potential to extend MENT to six or more dimensions. The problem is then to tune the sampling parameters such that each projection is a close approximation of the distribution function in Eq.~\eqref{eq:ment_solution}. The distribution function may be expensive to evaluate because it requires $K$ transformations of the coordinates through the particle tracking model. To increase the sampling speed, we therefore run multiple chains in parallel, evolving a small set of points through the phase space rather than a single point. We evolve the points until the chains ``mix'', i.e., until they are indistinguishable.

\section{Demonstrations}\label{sec:method-demonstration}

The rest of this paper demonstrates particle-based 6D MENT in two examples. In the first example, we use MCMC sampling to reconstruct a 6D Gaussian Mixture (GM) distribution from its 2D marginals, showing that the MCMC approach is stable in at least one case with a known forward model and multimodal ground truth distribution. In the second example, we deploy the same algorithm to data from the Argonne Wakefield Accelerator (AWA).

Reconstructions are simplified if the scale and linear correlations between variables are known. The covariance matrix $\bm{\Sigma} = \langle \bm{x}\bm{x}^T \rangle$ provides this information. In the following examples, we assume the $2 \times 2$ block-diagonal elements of the covariance matrix are available from previous measurements and reconstruct the distribution in normalized coordinates $\bm{z} = \bm{V}^{-1} \bm{x}$, where $\bm{\Sigma} = \bm{V} \bm{V}^T$ and $\bm{V}$ is a symplectic matrix \footnote{One way to fit the 4D or 6D covariance matrix to all measured projections is to leverage the Cholesky decomposition $\bm{\Sigma} = \bm{L}\bm{L}^T$, where $\bm{L}$ is a lower-diagonal matrix with positive diagonal elements, which ensures the covariance matrix is positive semi-definite. Alternatively, the diagonal or block-diagonal elements of the covariance matrix are typically available from faster measurements of the 2D phase space distributions $\{\rho(x, x'), \rho(y, y'), \rho(z, z') \}$. Cross-plane correlations ignored in this scenario are typically much weaker than within-plane correlations from, i.e., quadrupole focusing. If 2D measurements are unavailable, one may estimate the beam size by sampling from a uniform MENT prior and refining until the beam size can be estimated. (The prior is nonzero only in the regions allowed by the measurements.)}.

Our first example uses a 6D Gaussian Mixture (GM) distribution as the ground truth. The distribution is the superposition of seven Gaussian distributions, each with a random mean and variance;  it is a good candidate for initial testing because it is multimodal, easy to sample from, and has clear structure when projected onto lower dimensions. We collected the 2D marginal projections of the distribution as training data \footnote{For a fixed number of measurements $K = N (N - 1) / 2$ of a distribution without linear correlations between planes, we hypothesize that the ideal set of measurements is the set of $K$ two-dimensional marginals.}. We used the Metropolis-Hastings (MH) algorithm to sample from the 6D MENT distribution defined by these projections. We used a Gaussian prior with covariance matrix $\bm{\Sigma}$, where $\bm{\Sigma}$ is the assumed covariance matrix described above. We used a Gaussian jumping distribution with covariance matrix $0.25 \hat{\bm{\Sigma}}$. We simulated the projections on each iteration by running 500 parallel MCMC chains for 1000 steps each to generate $5 \times 10^5$ total samples. The starting point of each chain was sampled from a Gaussian with covariance matrix $0.25 \hat{\bm{\Sigma}}$. Gaussian filters were used to remove statistical noise from the binning process. 

\begin{figure*}
    \centering
    \includegraphics[width=\linewidth]{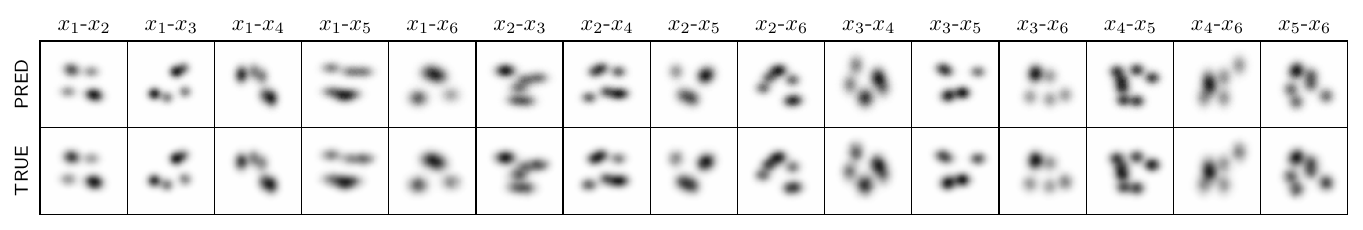}
    \caption{Two-dimensional marginal projections of Gaussian Mixture distribution. Top: MENT prediction; bottom: ground truth.}
    \label{fig:gmm}
\end{figure*}

The MENT algorithm converged in two iterations, with each iteration taking less than one minute to complete. Fig.~\ref{fig:gmm} shows that the 2D projections of the reconstructed and true distributions are identical.

Our second example applies the same algorithm to the open-source data set and accelerator model described in \cite{Roussel_2024_6d}. The measurements include a set of $x$-$y$ beam images after a quadrupole magnet, transverse deflecting cavity (TDC), and dipole bend at the Argonne Wakefield Accelerator (AWA). In \cite{Roussel_2024_6d}, the Generative Phase Space Reconstruction (GPSR) algorithm was used to fit a 6D phase space distribution to 20 images, with each image corresponding to a different quadrupole, TDC, and dipole field strength. The accelerator was modeled in Bmad-X, a differentiable particle tracking code. 

We assumed the block-diagonal elements of the covariance matrix were known from previous measurements \footnote{In this case, we extracted the covariance matrix elements from the previous GPSR fit.}. We ran MENT for 5 iterations with learning rate $\omega = 0.25$, sampling $5 \times 10^5$ particles per iteration with 1000 parallel chains. With these parameters, each iteration took approximately 8 minutes on a 2021 MacBook Pro computer with an Apple M1 Max chip. (Tracking was performed on a CPU, not a GPU.)

\begin{figure*}
    \centering
    \includegraphics[width=\linewidth]{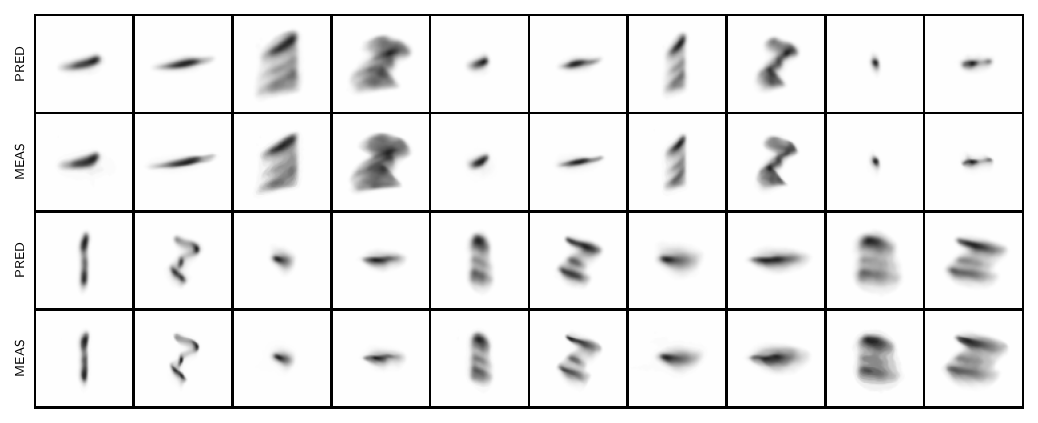}
    \caption{6D MENT applied to the Argonne Wakefield Accelerator (AWA) data set \cite{Roussel_2024_6d}. Rows 1/3 show simulated images and rows 2/4 show measured images.}
    \label{fig:awa-proj}
\end{figure*}

Fig.~\ref{fig:awa-proj} shows the measured and simulated images. Reasonable agreement is found after one or two iterations, with closer agreement after five iterations.
Reconstructed features are similar to the GPSR reconstruction in \cite{Roussel_2024_6d}, indicating the problem is well-constrained. Two-dimensional views of the reconstructed distributions are included in the Supplemental Material \footnote{See Supplemental Material at [URL will be inserted by publisher] for 1D and 2D projections of the reconstructed 6D phase space distribution.}. As a brief comparison, we sampled $9 \times 10^4$ particles from the MENT and GPSR models. For the training images, the average fractional error between measured and predicted pixels was 1.06\% for GPSR and 0.82\% for MENT; for the testing images, the error wass 0.93\% for GPSR and 0.72\% for MENT. We also estimated the entropy of each distribution by training a normalizing flow \cite{Durkan_2019} on each set of 6D coordinates. The flow-based estimate gives an absolute entropy of 6.58 for GPSR and a slightly higher 7.61 for MENT. Although the MCMC sampling parameters could be tuned more methodically, we conclude that MENT has provided a plausible solution.

\section{Conclusion}\label{sec:conclusion}

We have proposed a modified maximum-entropy (MENT) algorithm that scales to six-dimensional phase space. The method uses particle sampling, rather than numerical integration, to compute projections of the MENT distribution function. We showed that MCMC algorithms, such as the Metropolis-Hastings (MH) algorithm, can be used for six-dimensional reconstructions.

Particle-based MENT may be useful for accelerator physics research. MENT finds the most conservative distribution compatible with the data, making it a valuable tool when evaluating experimental results or comparing to other reconstruction algorithms. Interaction with trusted particle tracking codes is straightforward, as the only restriction on the tracking code is that particles must not interact with each other. Furthermore, there are no restrictions on the types of distributions MENT can represent. Possible applications and research topics include reconstructions in nonlinear lattices, reconstructions with self-consistent space charge forces, and high-dimensional and high-dynamic-range reconstructions \cite{Hoover_2023}. 

Algorithmic improvements may involve tuning of the MCMC parameters \footnote{The distribution exhibits some fluctuations during training which may indicate that the sampling chains have not fully converged to the target distribution. However, the final distribution seems fairly stable with respect to random number generator seed. Additional convergence diagnostics (beyond visual checks) should be used in future studies to resolve this issue.}, investigation of alternative sampling methods such as dynamic nested sampling \cite{Higson_2019} or flow-based generative models \cite{Midgley_2022}, and the use of parallel computing to speed up the sampling process. Additionally, a major limitation of MENT is its lack of uncertainty quantification. MENT selects a single distribution $\rho(\bm{x})$---the simplest distribution allowed by the constraints---but does not consider the plausibility of other nearby distributions. Consider a vector $\bm{\theta}$ parameterizing the distribution $\rho(\bm{x} | \bm{\theta})$. The points $\left \{ \bm{\theta} \right \}$ compatible with the constraints form a \textit{statistical manifold} \cite{Caticha_2021}. We are ultimately interested in a probability distribution over $\bm{\theta}$, conditional on the measurements. Evaluating the shape of this higher-level distribution in the maximum-entropy framework is left as future work.

\section{Acknowledgements}

I thank J. C. Wong for fruitful discussions on maximum-entropy methods and R. Roussel for supplying the AWA data and GPSR model. This manuscript has been authored by UT Battelle, LLC under Contract No. DE-AC05-00OR22725 with the U.S. Department of Energy. The United States Government retains and the publisher, by accepting the article for publication, acknowledges that the United States Government retains a non-exclusive, paid-up, irrevocable, world-wide license to publish or reproduce the published form of this manuscript, or allow others to do so, for United States Government purposes. The Department of Energy will provide public access to these results of federally sponsored research in accordance with the DOE Public Access Plan (http://energy.gov/downloads/doe-public-access-plan).

\section{Data and code availability}

The data and code used in this paper are available online \footnote{{https://doi.org/10.5281/zenodo.15412851}}.

\bibliography{main}

\end{document}